\documentclass{aa}  

\usepackage[varg]{txfonts}

\usepackage{txfonts}
\usepackage{lipsum}
\usepackage{subcaption}         % necessary for continued figures, example in section 3
\usepackage{lscape}             % to rotate a single page table, example in appendix.
\usepackage{placeins}           % useful with \FloatBarrier, to keep 
\usepackage{float}

\begin{document} 

\title{Nongravitational Motion of 3I/ATLAS Far from the Sun: Robust Detection, Weak Constraints on the Activity Law}

\titlerunning{Nongravitational Motion of 3I Far from the Sun}

   \author{Małgorzata Królikowska\inst{1}\corrauth{mkr@cbk.waw.pl}       % use \corrauth for the corresponding author
   	\and Federico Spada\inst{2}\email{federico.spada@gmail.com} \and Luke Dones\inst{3}\email{henry.dones@swri.org}
%	\and F. Déliat\inst{1}\email{fab@nestor-edp.org}\fnmsep\thanks{NASA fellow (shows the usage of elements in the author field)}
%	\and C. Brassac\inst{1}\email{cat@nestor-edp.org}
}

 \institute{Centrum Badań Kosmicznych Polskiej Akademii Nauk (CBK PAN), Bartycka 18A, 00-716 Warszawa, Poland \and 
Università di Catania, Dipartimento di Fisica e Astronomia, Sezione Astrofisica, Via S. Sofia 78, 95123 Catania, Italy  \and 
Department of Space Studies, Southwest Research Institute, 1301 Walnut St Suite 400, Boulder, CO 80302, USA}
  
%   \and Institute of Physics, Jan Kochanowski University, Uniwersytecka 7, 25-406 Kielce, Poland}

%%%%%%%%%%%%%%%%%%%	\and
%%%%%%%%%%%%%%%%%%%		Centrum Badań Kosmicznych Polskiej Akademii Nauk (CBK PAN),
%%%%%%%%%%%%%%%%%%%		Bartycka 18A, 00-716 Warszawa, Poland}
%   Space Research Centre of Polish Academy of Sciences, Bartycka 18A, PL00-716, Warszawa, Poland}
%Space Research Centre of the Polish Academy of Sciences, Bartycka 18A, 00-716 Warsaw, Poland}

\date{Received XXXXXX; accepted XXXXX}

% \abstract{}{}{}{}{} 
% 5 {} token are mandatory

\abstract
% context heading (optional)
{3I/ATLAS provides an unprecedented astrometric data set for studying nongravitational (NG) effects in an interstellar object, but existing orbital solutions depend strongly on assumptions regarding the heliocentric activity law and the treatment of the astrometric data.}
% aims heading (mandatory)
{We investigate the extent to which astrometric observations constrain the heliocentric dependence of the NG activity function, focusing on the identifiability of different functional forms of $g(r)$ within standard NG orbit-determination frameworks.}
% methods heading (mandatory)
{Astrometric observations obtained at heliocentric distances larger than 3.49 au before perihelion and 2.67 au after perihelion were analysed. Purely gravitational and nongravitational orbital solutions were determined using several activity laws, including the standard water-ice sublimation law and flatter more volatile ice prescriptions. The resulting residual distributions were evaluated using normality tests and analyses of temporal trends in the observed-minus-calculated residuals.}
% results heading (mandatory)
{The distant pre-perihelion observations are well described by a purely gravitational (GR) orbit, whereas the distant post-perihelion data exhibit a statistically significant non-Gaussian residual distribution under the GR model. When the distant pre- and post-perihelion observations are combined, nongravitational acceleration becomes clearly detectable through a significant reduction in the residual rms and the disappearance of systematic residual trends. However, the astrometric data do not uniquely distinguish between the tested activity laws: standard water-ice, more volatile and extreme volatile prescriptions produce statistically comparable orbital fits despite their markedly different heliocentric dependences.}
% conclusions heading (optional)
{The existence of nongravitational acceleration is therefore considerably better constrained by astrometry than its physical origin or heliocentric dependence.}


\maketitle

\nolinenumbers

\section{Introduction}\label{sec:intr}

3I/ATLAS is the first interstellar object with such a large astrometric data set available for testing different scenarios of non-gravitational (NG) acceleration in its motion. In the Minor Planet Center\footnote{\tt https://minorplanetcenter.net/db\_search} database, 8498 observations over the interval 2025 May 08 -- 2026 April 14 are publicly available as of October 2026. These measurements span heliocentric distances from 6.1 au before perihelion, through perihelion passage at 1.356 au, to 6.1 au after perihelion.

Before discussing the identifiability of the activity function from astrometric data, it is useful to summarize existing NG orbital solutions for 3I/ATLAS.
%, including both operational ephemeris solutions and dedicated dynamical studies. 
It is important to emphasize that these solutions are not purely observational products, but arise from explicit orbit determination procedures in which different assumptions about the heliocentric activity function $g(r)$, data weighting, and dynamical configuration lead to distinct realizations of the inferred NG parameters.

The baseline reference is the solution provided by the JPL Small-Body Database Browser\footnote{\tt https://ssd.jpl.nasa.gov/tools/sbdb\_lookup.html\#/}, which incorporates a fixed functional form of the heliocentric activity law $g(r)$ (typically described there as water-ice sublimation, CO sublimation, or a CO$_2$-like approximation depending on the solution version) and estimates the corresponding Marsden-type NG parameters \citep{marsden-sek-ye:1973}, in many cases including the asymmetric time of maximum outgassing (see Sect.~\ref{subsec:models}). For 3I/ATLAS,
JPL uses an asymmetric activity law described there as `Nongravitational accels. using g(r)= (1 au/r)$^2$ for CO$_2$'. This solution serves as the standard model against which more detailed NG analyses are compared. 

\begin{table*}[h!]
	\setlength{\tabcolsep}{2.0pt}  	\centering
	\small
	\begin{tabular}{lccccccc}
		\hline
		Solution & Data arc & N obs & $A_1$ & $A_2$ & $A_3$ & $DT$ (days) & RMS  \\
		\hline
		
		JPL SBDB \\ (NG solution \#54) 
		& 2025 May 15 -- 2026 Feb 19  
		& 782  
		& 5.320$\pm$0.122 
		& 1.148$\pm$0.214
		& $-0.685\pm 0.018$ 
		& $+9.48 \pm 1.43$ 
		& $0.425$  \\
 	
	\cite{Eubanks2025} \\  E-data 
		& 2025 May--Dec 1
		& $\sim$4200
		& $-2.282\pm 0.455$ 
		& $9.221\pm 0.632$ 
		& $-0.411\pm 0.022$ 
		& $-34.6 \pm 2.6$ 
		& $>1.0$ \\
 		
	\cite{Spada2026} \\ 'ng2' solution 
		& 2025 May 8 -- 2026 Feb 19 
		& $\sim$8000 
		& $5.381\pm 0.085$
		& $1.149\pm 0.139$ 
		& $-0.760\pm 0.010$ 
		& $+7.49\pm 0.92$  
		& $0.77$ \\ \\
		
	\cite{Thoss2026} 
+		& JPL-based NG solution  & inherited & not refit & not refit & not refit & not fitted         & -- \\   
(interpretation only) 		& (version-dependent) 	 &           &           &           &           &   (uses JPL input) &    \\
		\hline
	\end{tabular}
	\caption{NG orbital solutions for 3I/ATLAS; A$_1$, A$_2$, and A$_3$ parameters are in units of $10^{-8}$au day$^{-2}$, and $DT$ in days; RMS is dimensionless for the JPL solutions and RMS values from other sources are given in arcsec. The JPL solution represents the current orbital solution \#54 (as October 12, 2026). The analysis of \cite{Thoss2026} is based on a previous JPL NG solution version and therefore depends on a different parameter realization of the same model class. }\label{tab:other-solutions}
\end{table*}

%A1 =     5.381  ±  0.085   (10^-8 au/d^2)  < ng2 Spada et al
%A2 =     1.149  ±  0.139   (10^-8 au/d^2)
%A3 =    -0.7597 ±  0.00977 (10^-8 au/d^2)
%DT =     7.494  ±  0.918   (days)

Independent investigations, such as \cite{Eubanks2025} and \cite{Spada2026}, performed independent NG orbit determinations from astrometry, allowing for additional degrees of freedom in the activity model, including asymmetric, time-shifted forms of $g(r)$ characterized by the parameter $DT$  \citep{yeomans-chodas:1989,sitarski:1994b}. Moreover, \cite{Spada2026} considered a more general class of asymmetry by allowing different pre- and post-perihelion activity slopes. Both studies reported statistically significant NG signals and obtained well-determined NG parameters, including $DT$. However, the derived values of $DT$ differ substantially and even have opposite signs. The most plausible source of this discrepancy is the different observational arc adopted in each study, while differences in weighting and data-processing procedures are likely of secondary importance.

Importantly, the choice of CO/CO$_2$-like activity laws used in several NG models is not ad hoc, but motivated by independent observational evidence for volatile composition in cometary and interstellar objects, where CO$_2$ and CO are observed to be major volatile species at heliocentric distances where water-ice sublimation becomes inefficient. Independent spectroscopic and compositional studies provide strong evidence that CO$_2$ is a major volatile species in the coma of 3I/ATLAS. JWST/NIRSpec observations reported by \cite{Cordiner2025JWST3IATLAS} revealed a CO$_2$-rich coma at heliocentric distances of about 3.3 au, with a CO$_2$/H$_2$O abundance ratio among the highest measured in cometary environments. Similar conclusions were reached from SPHEREx observations \citep{ Lisse_SPHEREx_2026b}, which indicate substantial contributions from both CO$_2$ and CO to the observed activity.

Complementary modelling by \cite{Maggiolo2025GCR3IATLAS} suggests that enhanced CO$_2$ abundances may result from irradiation-driven processing of CO-rich ices in the nucleus. Taken together, these studies provide a strong physical motivation for considering CO$_2$-like activity laws in NG orbit determination. However, they do not by themselves determine the functional form of the heliocentric activity law.

A different approach was adopted by \cite{Thoss2026}, who did not determine NG parameters directly from astrometric observations but instead examined the physical interpretation of previously derived NG solutions. Using the JPL orbital solution as input, they compared the implications of standard water-ice sublimation and a CO$_2$-driven activity law and concluded that the latter provides a more plausible physical explanation of the inferred acceleration. Their study therefore addresses the physical interpretation of an already determined NG solution rather than providing an independent astrometric constraint on the activity law itself.

A complementary perspective was presented by \cite{Cloete2025}, who examined the extent to which the available astrometry is capable of constraining NG accelerations in 3I/ATLAS. Their analysis showed that the detectability of a NG signal depends sensitively on data weighting and modelling assumptions, emphasizing that astrometric evidence alone may not uniquely determine the preferred dynamical model.

These considerations motivate a broader question regarding the information content of astrometric observations. The purpose of this work is therefore not only to investigate whether NG~acceleration remains detectable when only distant observations are considered, but also to examine how strongly the available astrometry constrains the heliocentric dependence of that acceleration. Particular attention is paid to the extent to which different activity laws can be distinguished on the basis of orbit determination solutions derived from the same astrometric data set.

Why do we restrict the analysis to distant observations? Two independent arguments motivate this choice. First, the detectability of NG effects in the distant astrometry of 3I/ATLAS has not yet been established. Second, the current understanding of the comet's volatile evolution indicates that different sublimation regimes dominate at different heliocentric distances.

The available observations indicate that 3I/ATLAS underwent a transition between different sublimation regimes during its passage through the inner Solar System. At large heliocentric distances before perihelion (approximately 3.3 au), JWST/NIRSpec observations revealed strongly enhanced CO$_2$ activity relative to H$_2$O, indicating that CO$_2$ sublimation appears to dominate the activity at larger pre-perihelion heliocentric distances \citep{Juncen_Li:2026,Lisse_SPHEREx_2026b}. Closer to perihelion, observations obtained at heliocentric distances of $r_h \simeq 1.36$--$1.37$ au with the IRAM 30-m telescope and at $r_h = 1.36$--$1.68$ au with JUICE/MAJIS showed an active coma containing H$_2$O, CO$_2$, and CO, with water sublimation becoming an increasingly important contributor to the cometary activity near perihelion \citep{Biver2026,BockeleeMorvan2026}. After perihelion, JWST/MIRI observations at $r_h = 2.20$ and $2.54$ au showed that the H$_2$O production declined more rapidly than that of the more volatile species, resulting in an increasing relative importance of CO and a persistently high CO$_2$/H$_2$O ratio \citep[][and references therein; see their Fig.~5]{Belyakov2026}. Overall, these observations support a scenario in which CO$_2$ is the principal driver of activity at pre-perihelion larger heliocentric distances, water sublimation becomes increasingly important near perihelion, and the relative contribution of CO increases during the post-perihelion evolution.

This evolutionary picture suggests that modelling the entire currently available astrometric data set, spanning heliocentric distances from 6.1 au before perihelion, through perihelion at 1.35 au, to 6.1 au after perihelion, with a single NG~law based on CO$_2$ sublimation over the entire observed heliocentric distance range would represent a simplification of the underlying physical picture. The changing volatile inventory implies that the functional form of the NG acceleration is expected to evolve along the orbit rather than remain adequately described by a single sublimation law over the full observed heliocentric distance range. 

The remainder of this paper is organized as follows. Section~\ref{sec:material_methods} describes the astrometric data sets, the orbit determination procedure, the tested dynamical models, and the statistical methods used to assess the residual distributions. Sections~\ref{sec:pre} and \ref{sec:post} analyze the distant pre- and post-perihelion data arcs separately. The results obtained from the combined pre- and post-perihelion data set are presented in Section~\ref{sec:full_models}, including both symmetric and asymmetric NG solutions. Section~\ref{sec:other_ng_models} examines the extent to which the available astrometry constrains the adopted activity law.  The results are compared with those obtained using alternative post-perihelion data selections in Section~\ref{sec:test}. The principal conclusions are summarized in Section~\ref{sec:conclusions}.

\section{Astrometric Material and Orbit Determination} \label{sec:material_methods}

In this study, the observations obtained closest to perihelion were deliberately excluded from the analysis. Consequently, the orbital solutions presented below are based exclusively on measurements acquired outside the heliocentric distance interval extending from 3.49~au before perihelion, through perihelion at 1.356~au, to 2.67~au after perihelion. This excluded interval contains approximately 4000 astrometric observations and corresponds to the region where NG~effects are expected to be strongest.

\begin{table}[ht]
	\caption{Astrometric data sets used in this study; rms values are given for the purely gravitational orbital solutions.}
	\label{tab:data-sets}
	\setlength{\tabcolsep}{1.0pt}    \centering
	\begin{tabular}{cccccc}
		\hline \hline
		set name   & data arc & Number        & heliocentric  & rms      \\
		&          & of obs.       & distances [au] & [arcsec] \\
		\hline
		data-set1   & 2025 05 15--2025 07 31  & 2646  & 6.08 -- 3.49    & 0.2558  \\
		data-set2   & 2026 01 01--2026 04 14  & 1757  & 2.67 -- 6.09    & 0.2496  \\
		data-set3   & 2025 05 15--2026 04 14  & 4403  & as in data-set1 & 0.3286  \\
		&                              &       & \& data-set2    &       \\
		\hline
	\end{tabular}
\end{table}

Three data sets were constructed and subsequently analysed. The first two, data-set1 and data-set2, consist of the distant pre-perihelion and post-perihelion observational arcs, respectively. Their basic characteristics are summarized in Table~\ref{tab:data-sets}. For both data sets, selection and weighting procedures were applied during the determination of purely gravitational orbital solutions.

The model of motion and the orbit determination procedure adopted in this study are described in detail by \citet[][and references therein]{kroli_dyb:2017}. The basic dynamical model, including the perturbations from all planets and relativistic effects, was kept fixed for all solutions, ensuring a consistent comparison of the different NG models. The formalism used to model nongravitational (NG) acceleration is outlined below. The weighting method was originally introduced by \citet{bielicki-sitarski:1991} and later discussed in detail by \citet{krolikowska-sit-soltan:2009}. The procedure involves two distinct stages. First, the astrometric data are subjected to statistical filtering to identify and remove outlying observations in right ascension and declination. This initial filtering takes the reported measurement uncertainties into account through the adopted statistical rejection criteria. The remaining observations are then assigned relative weights according to the actual scatter of their residuals. These weights are determined iteratively together with the orbital solution, so that the orbital parameters and the observational weights are updated until convergence.

In this way, observations from stations showing higher internal consistency receive greater statistical weight, while observations characterized by larger scatter contribute less to the final solution. Because the weights are derived from the residual statistics and are updated together with the orbital solution, the final weighting is intentionally dependent on the orbit being determined. This provides a self-consistent framework for the heterogeneous astrometric data set and avoids relying exclusively on a priori uncertainty classes assigned to individual observations. The procedure therefore represents an alternative to weighting schemes based primarily on reported measurement uncertainties, while retaining their use at the initial statistical filtering stage. As an additional check, we verified that the observations from stations generally regarded as providing high-precision astrometry receive relative weights greater than 1 in our procedure. This provides an additional indication that the adopted weighting scheme does not systematically downweight observations from the potentially most precise observing stations.

In data-set3, the pre- and post-perihelion observations were combined into a single data set comprising 4403 astrometric measurements obtained between 2025 May 15 and 2026 April 14, while retaining the weights derived from the independent processing of the two arcs (Table~\ref{tab:data-sets}). The resulting observational arc spans nearly eleven months and covers large heliocentric distances on both sides of perihelion. It is important to emphasize that the selection and weighting procedures were carried out independently for the pre- and post-perihelion observations before the two data sets were merged.

\subsection{Tested Dynamical Models}
\label{subsec:models}

\begin{figure*}
	\centering
	\includegraphics[width=1.00\linewidth]{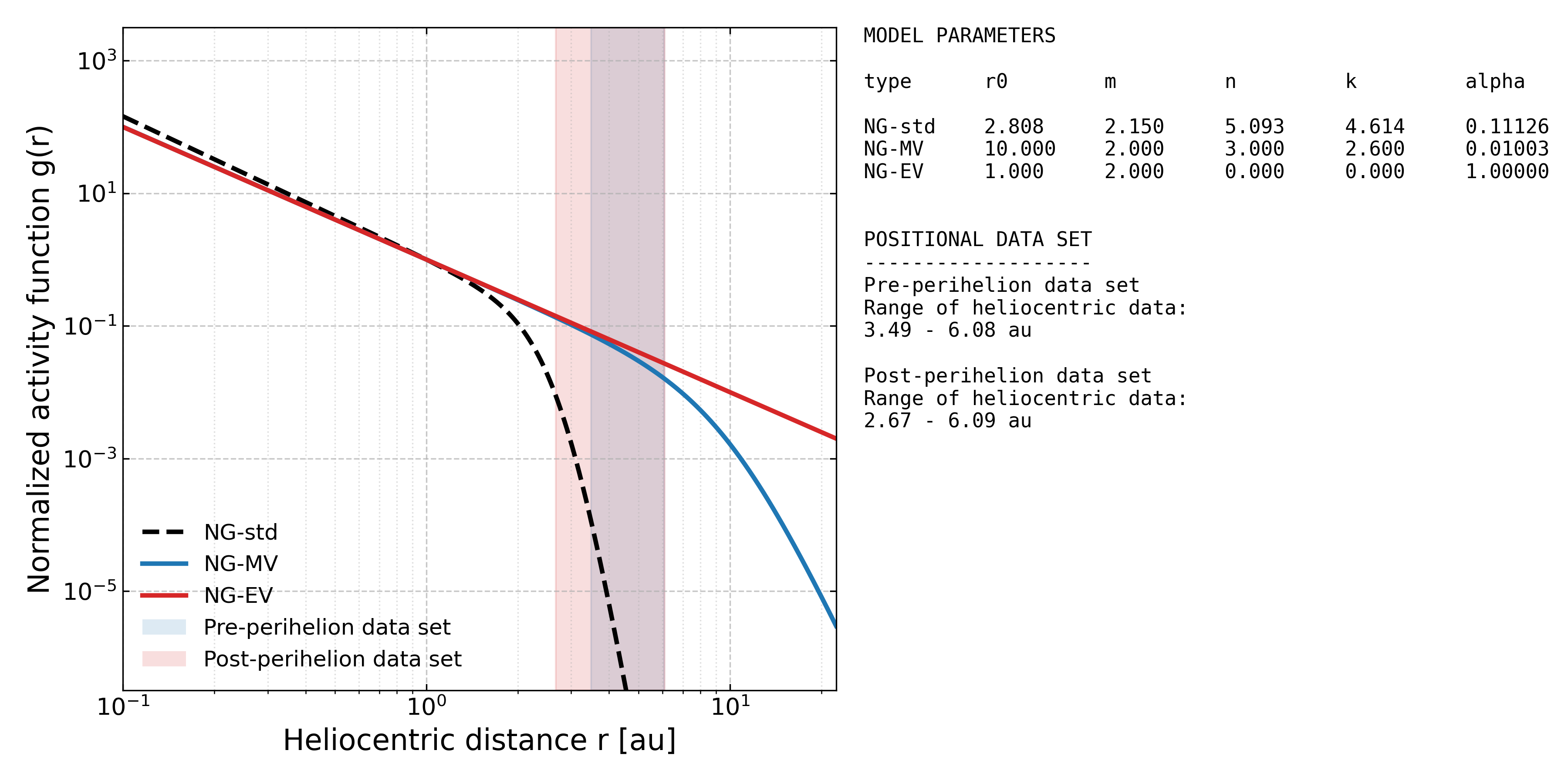}
	\caption{Three $g(r)$-like functions tested in this study. The ranges of the pre- and post-perihelion positional observations (data-set1 and data-set2, respectively; Table~\ref{tab:data-sets}) are indicated by the blue  and red vertical bands, respectively.}
	\label{fig:g_r_plot}
\end{figure*}

Four dynamical models were tested for each of the three data sets defined above: a purely gravitational model (GR) and three nongravitational (NG) models based on different $g(r)$-like functions. These include the standard water-ice sublimation law (NG-std) and two alternative prescriptions corresponding to the sublimation of more volatile species (NG-MV model) and extremely volatile species (NG-EV model).

The functional form of $g(r)$ depends not only on the volatile species responsible for the outgassing, but also on the adopted sublimation model of the cometary nucleus, for example whether sublimation is assumed to
occur under isothermal conditions, exclusively from the subsolar point, or according to an intermediate and more realistic thermal model. Consequently, different physical assumptions may lead to somewhat
different heliocentric dependencies even for the same volatile species. For this reason, the NG-MV and NG-EV prescriptions adopted here are intended to represent generic classes of activity laws associated with more and extremely volatile materials, rather than unique sublimation models of specific species such as CO$_2$ or CO. Nevertheless, these two volatiles provide the primary physical motivation for considering such
activity laws, as independent observations have demonstrated that both CO$_2$ and CO contributed substantially to the activity of 3I/ATLAS at large heliocentric distances.

To determine NG orbits, we adopted the standard formalism proposed by \citet{marsden-sek-ye:1973}, in which the three components of the NG acceleration acting on a comet are assumed to be proportional to a symmetric $g(r)$ function,

\begin{eqnarray}
F_{i}=A_{\rm i} \> g(r),& A_{\rm i}={\rm ~constant~~for}\quad{\rm i}=1,2,3,\nonumber\\
& \quad g(r)=\alpha(r/r_{0})^{-m}[1+(r/r_{0})^{n}]^{-k},
\label{eq:g_r}
\end{eqnarray}

\noindent where $F_{1},\,F_{2},\,F_{3}$ denote the radial, transverse, and normal components of the NG acceleration, respectively, and the radial component is defined as positive outward along the Sun--comet line.

Three variants of the $g(r)$ function were considered in this study. Their heliocentric-distance dependences, together with the adopted values of the parameters $m$, $n$, $k$, $r_0$, and $\alpha$, are shown in Fig.~\ref{fig:g_r_plot}. The normalization constant $\alpha$ is determined from the condition $g(1\,{\rm au})=1$.

A clear message emerges from Fig.~\ref{fig:g_r_plot}. In principle, within the heliocentric distance range covered by the available astrometric observations, the classical water-ice sublimation law is readily distinguishable from the two alternative prescriptions. In contrast, the MV-driven (blue curve) and EV-driven (red curve) models exhibit very similar heliocentric dependences throughout the observational interval marked by the shaded bands. Consequently, the ability of the available astrometric data to discriminate between these two activity laws and to identify one of them as a more realistic description of the cometary activity is questionable from the outset.

This immediately suggests that a meaningful comparison between the NG-MV and NG-EV models would require observations extending to substantially larger heliocentric distances than those currently available. In practice, observations reaching distances of the order of $\sim20$~au would be desirable, where the corresponding $g(r)$-like functions begin to diverge significantly. As will be shown in the following sections, the results fully support this expectation.

For data-set3, we additionally attempted to determine asymmetric NG models for all three activity laws. In these models, the symmetric function $g(r(t))$ is replaced by the asymmetric form $g(r(t-DT)$ \citep{yeomans-chodas:1989,sitarski:1994b}. In this case, four NG parameters are determined simultaneously with the orbital elements, namely $A_1$, $A_2$, $A_3$, and $DT$, where $DT$ represents the temporal displacement of the maximum of the $g(r)$ function relative to the time of perihelion passage.

\subsection{Statistical tests}
\label{subsec}

Like \citet[][]{kroli-dyb:2010}, we consider the detection of NG~effects to be justified when the following indicators are present:

\noindent (i) a reduction in the root mean square (rms) of the residuals for the NG solution relative to the purely gravitational (GR) solution;

\noindent (ii) a reduction, or preferably the disappearance, of systematic trends in the temporal distribution of the Observed--Calculated (O--C) residuals in right ascension and/or declination, provided that such trends are present in the corresponding GR solution;

\noindent (iii) an O--C residual distribution that is closer to Gaussian for the NG solution than for the GR solution.

To evaluate whether the residual distributions are consistent with a normal distribution -- criterion (iii) -- we applied two complementary statistical tests using the \texttt{anderson} and \texttt{normaltest} functions, respectively, from the \textit{scipy.stats} module of the SciPy Python library \citep{Virtanen2020}. The Anderson--Darling (A--D) test \citep{Anderson1954} is particularly sensitive to deviations in the tails of the distribution, whereas the D'Agostino--Pearson test \citep{DAgostino1973,DAgostino1990} combines measures of skewness and kurtosis into a single test statistic, thereby providing a complementary assessment of departures from normality. The combined use of these tests provides a robust assessment of the Gaussian character of the astrometric residuals. The Anderson--Darling test has also been applied in astrophysical studies to assess whether observational distributions are consistent with Gaussian expectations, for example in analyses of flux distributions and variability properties of astrophysical sources \citep{Shah2020}.

For the D'Agostino--Pearson test, the null hypothesis of normality was rejected when the corresponding p-value was smaller than 0.05. For the Anderson--Darling test, the null hypothesis was rejected when the test statistic exceeded the critical value corresponding to the 5\% significance level. The latter criterion follows the implementation of the Anderson--Darling test in {\it scipy.stats}, which provides critical values rather than a direct p-value estimate. 

% For Luke:  I think there may be a small misunderstanding here. For the D'Agostino--Pearson test I provide p-values, separately for the RA and Dec residuals. However, the SciPy implementation of the Anderson--Darling test (scipy.stats.anderson) does not return p-values; it provides the test statistic together with critical values for different significance levels. Therefore, for A-D I report the statistic and apply the 5% critical value criterion. I will add a clarification in the text and in the table caption to make this explicit.

\section{Pre-perihelion data arc}
\label{sec:pre}

The pre-perihelion data arc extends from 2025 May 15 to 2025 July 31 and contains 2646 positional observations, including approximately 30 pre-discovery measurements. These pre-discovery observations significantly extend the observational arc and substantially improve the formal accuracy of the purely gravitational orbital solution. Nevertheless, the statistical properties of the residual distributions remain largely unchanged when comparing GR solutions with and without the pre-discovery data. It should be noted, however, that the pre-discovery observations comprise measurements from several facilities. During the selection and weighting procedure, almost all pre-discovery observations obtained by TESS (observatory code C57) were rejected as outliers, with only a single right ascension measurement formally retained. Since this isolated measurement carries essentially no statistical weight, it was ultimately excluded from the final orbital solution as well. During the selection and weighting procedure applied in the determination of the GR orbit, only about 1.5\% of the observations were rejected as obvious outliers.

The residuals in right ascension are fully consistent with a Gaussian distribution, yielding a fitted standard deviation of $\sigma_{\alpha}=0.256$~arcsec. The Anderson--Darling statistic is 0.70, below the 5\% critical value of 0.786, while the D'Agostino--Pearson test yields a $p$-value of 0.72. The skewness (0.04) and kurtosis (0.01) are both very close to zero, indicating an almost perfectly symmetric distribution with nearly Gaussian tails (see data-set1 in Table~\ref{tab:stat-pre-post}).

The declination residuals show somewhat larger departures from normality, with $\sigma_{\delta}=0.229$~arcsec. The Anderson--Darling statistic reaches 1.63, formally exceeding the critical value, whereas the D'Agostino--Pearson test does not reject normality ($p=0.10$). At the same time, the skewness (0.08) and kurtosis (0.12) remain small, suggesting that the deviations from a Gaussian distribution are weak. An additional test, performed after removing the outermost 2.5\% of the declination residuals (data-set1$_a$ in Table~\ref{tab:stat-pre-post}), reduces the Anderson--Darling statistic to 0.87, close to the critical value, while preserving a very small skewness (0.02) and only modest kurtosis ($-0.13$).

Overall, the residual distributions of the pre-perihelion observations are close to Gaussian and show no evidence of strong systematic effects. Combined with the absence of visible temporal trends in the O--C residuals, this indicates that the pre-perihelion arc is well represented by a purely gravitational orbit. No compelling evidence for additional dynamical effects is visible in the residuals. In addition, this solution may provide a useful starting point for studies of the comet's past dynamical evolution, particularly investigations of its trajectory before entering the Oort Cloud. For this reason, the corresponding orbital elements are listed in Appendix~\ref{sec:appA} (Table~\ref{tab:orbital_solutions}).

Accordingly, attempts to determine independent nongravitational solutions from this arc alone, using all three tested $g(r)$-like functions, do not yield statistically meaningful sets of NG parameters. This result becomes particularly interesting when compared with the 'ng2' solution derived by \citet{Spada2026} from a pre-perihelion data arc that includes observations obtained much closer to perihelion. In their solution, the radial parameter $A_1$ is determined with high significance, the normal component $A_3$ is also well constrained, whereas the transverse component $A_2$ remains the least certain. This comparison suggests that observations obtained at smaller heliocentric distances than those considered here contribute substantially to the successful determination of a nongravitational orbit.

\begin{table}[ht]
	\caption{ Results of the Anderson--Darling test (test statistic values) and the D'Agostino--Pearson normality test (p-values) for the right ascension and declination residuals. Subscripts denote progressively truncated residual distributions (see text for details).}  
	\label{tab:stat-pre-post}
	\setlength{\tabcolsep}{3.0pt}    \centering
	\begin{tabular}{cccccc}
		\hline \hline 
		set of   & model of & \multicolumn{2}{c}{A--D test} &  \multicolumn{2}{c}{D'A-P; p-value}  \\
		data arc & motion   &$\alpha$  & $\delta$         &   $\alpha$  &  $\delta$         \\    
		\hline  
		\multicolumn{6}{c}{Purely gravitational solutions} \\
		data-set1      & GR   & 0.696   & 1.62  & 0.720   & 0.102  \\
		data-set1$_a$  & GR   & --      & 0.87  &  --     & 0.343  \\
		data-set2      & GR   & 4.61    & 6.81  & 0.0076  & 0.0094  \\
		data-set2$_b$  & GR   & 3.94    & 6.72  & 0.0185  & 0.0124  \\
		data-set2$_c$  & GR   & 3.26    & 5.74  & 0.0025  & 0.122   \\
		data-set3      & GR   & 2.18    & 9.19  & 0.148   & 9$\cdot 10^{-9}$ \\ \\
	\hline  
		\multicolumn{6}{c}{Non-gravitational solutions} \\
		data-set3      & NG-std       & 2.49   & 6.64   & 0.420   & 0.0082 \\   
		data-set3      & NG-std-asymm & 2.33   & 6.08   & 0.466   & 0.0118 \\   
		data-set3      & NG-MV        & 2.54   & 6.24   & 0.207   & 0.0103 \\   
        data-set3      & NG-MV-asymm  & 2.28   & 5.84   & 0.443   & 0.0148 \\   
		data-set3      & NG-EV        & 2.55   & 6.19   & 0.208   & 0.0106 \\   
        data-set3      & NG-EV-asymm  & 2.29   & 5.83   & 0.451   & 0.0148 \\   
   \hline  
	\end{tabular}
\end{table}

\section{Post-perihelion data arc}
\label{sec:post}

The GR orbit fitted to the 1757 post-perihelion observations obtained between 2026 January 1 and 2026 April 14 yields an rms of 0.25 arcsec. During the selection and weighting procedure applied in the determination of the GR orbit, only about 1.8\% of the observations were rejected as obvious outliers. As in the pre-perihelion arc, no evident time-dependent trends are visible in the O$-$C residuals. However, the statistical properties of the residual distributions differ from those found before perihelion. 

For the nominal solution, both the right ascension and declination residuals deviate significantly from a Gaussian distribution, giving a formal fitted standard deviation of $\sigma_{\alpha}=0.269$~arcsec and $\sigma_{\delta}=0.203$~arcsec. The Anderson--Darling statistics are 4.61 and 6.81 for $\alpha$ and $\delta$, respectively, greatly exceeding the 5\% critical value of 0.785. The D'Agostino--Pearson test leads to the same conclusion, yielding p-values below 0.01 in both coordinates (Table~\ref{tab:stat-pre-post}). Thus, despite the satisfactory rms and the absence of visible temporal trends, the residual distributions exhibit statistically significant departures from normality.

To investigate whether these departures are caused primarily by a small number of extreme residuals in the distribution tails, the post-perihelion sample was re-analysed using progressively tighter residual limits. The purpose of this test was to determine whether removing the tails of the residual distributions would make them consistent with Gaussian distributions. In the first experiment (data-set2$_b$ in Table~\ref{tab:stat-pre-post}), residuals exceeding $\pm0.70''$ in right ascension and $\pm0.60''$ in declination were removed. This eliminated approximately 20 and 40 residuals from the corresponding distribution tails, respectively. In the second experiment (data-set2$_c$), even more restrictive limits of $\pm0.60''$ and $\pm0.55''$ were applied, removing about 60 residuals in $\alpha$ and 50 residuals in $\delta$. The results are summarized in Table~\ref{tab:stat-pre-post}.

Although truncation of the distribution tails decreases the residual dispersion and slightly lowers the Anderson--Darling statistic in right ascension, the residual distributions remain significantly non-Gaussian. In particular, the declination residuals show only modest improvement. Even after dozens of observations are removed, the Anderson--Darling statistic remains more than seven times larger than the critical value. The normality tests therefore indicate that the observed non-Gaussianity cannot be attributed solely to a small population of outliers located in the distribution tails.

This behaviour contrasts markedly with that observed in the pre-perihelion arc, where the residual distributions were close to Gaussian in both coordinates. The persistence of non-Gaussian residuals after perihelion suggests that an additional source of scatter affects a substantial fraction of the observations. Such behaviour may result from residual dynamical effects not fully represented by the adopted GR~model, systematic differences among astrometric data sets obtained at different observatories, or a combination of both effects.

The appearance of statistically significant non-Gaussian residuals only after perihelion is particularly noteworthy in the context of the subsequent nongravitational analysis. If an additional dynamical component is present, its astrometric signature appears to be concentrated predominantly in the post-perihelion observations rather than in the pre-perihelion arc, which exhibits nearly Gaussian residual statistics. This conclusion is influenced in part by the deliberate exclusion of observations obtained inside 3.49~au before perihelion. The adopted truncation was chosen for two reasons. First, it allows us to demonstrate that no detectable signature of NG~acceleration is present in the pre-perihelion positional data obtained between May and the end of July 2025. Second, it reduces the disparity between the numbers of observations available before and after perihelion while preserving the post-perihelion cutoff near 2.7--2.8~au, that is, close to the turnover distance of the classical water-ice sublimation law. This selection of observations highlights the markedly different behaviour of the comet before and after perihelion.

Attempts to determine independent nongravitational solutions from the post-perihelion arc alone, using all three tested $g(r)$-like functions, likewise do not yield statistically meaningful sets of NG parameters. This result should be understood as a lack of statistically significant NG~parameter determination from this relatively short data arc, rather than as evidence for the absence of NG~effects. The post-perihelion arc spans only about 3.5~months, compared with approximately 2.5~months for the pre-perihelion arc, whereas the combined data set provides a time baseline of nearly one year.

\begin{figure*}
	\centering
	\includegraphics[width=0.495\linewidth]{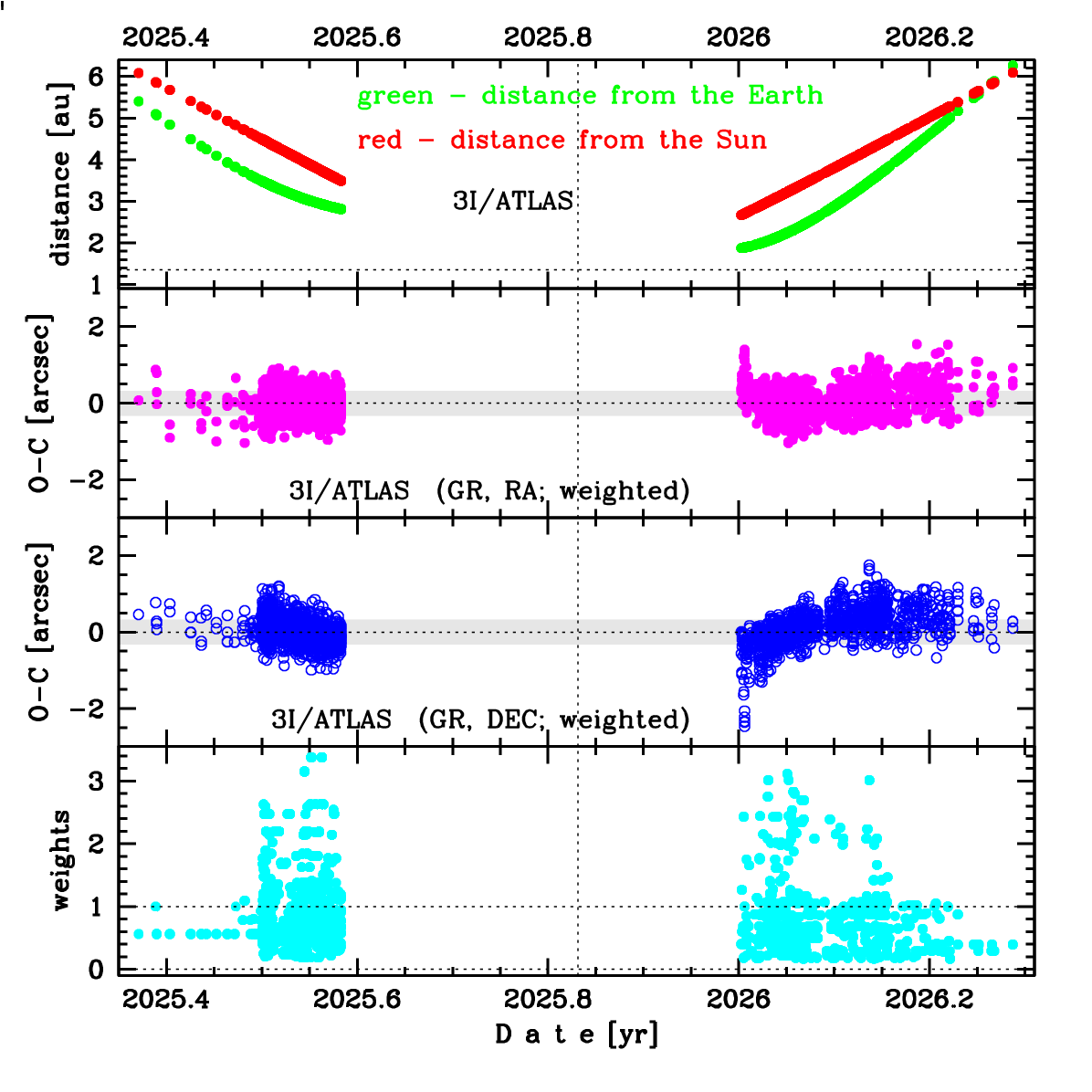}
	\includegraphics[width=0.495\linewidth]{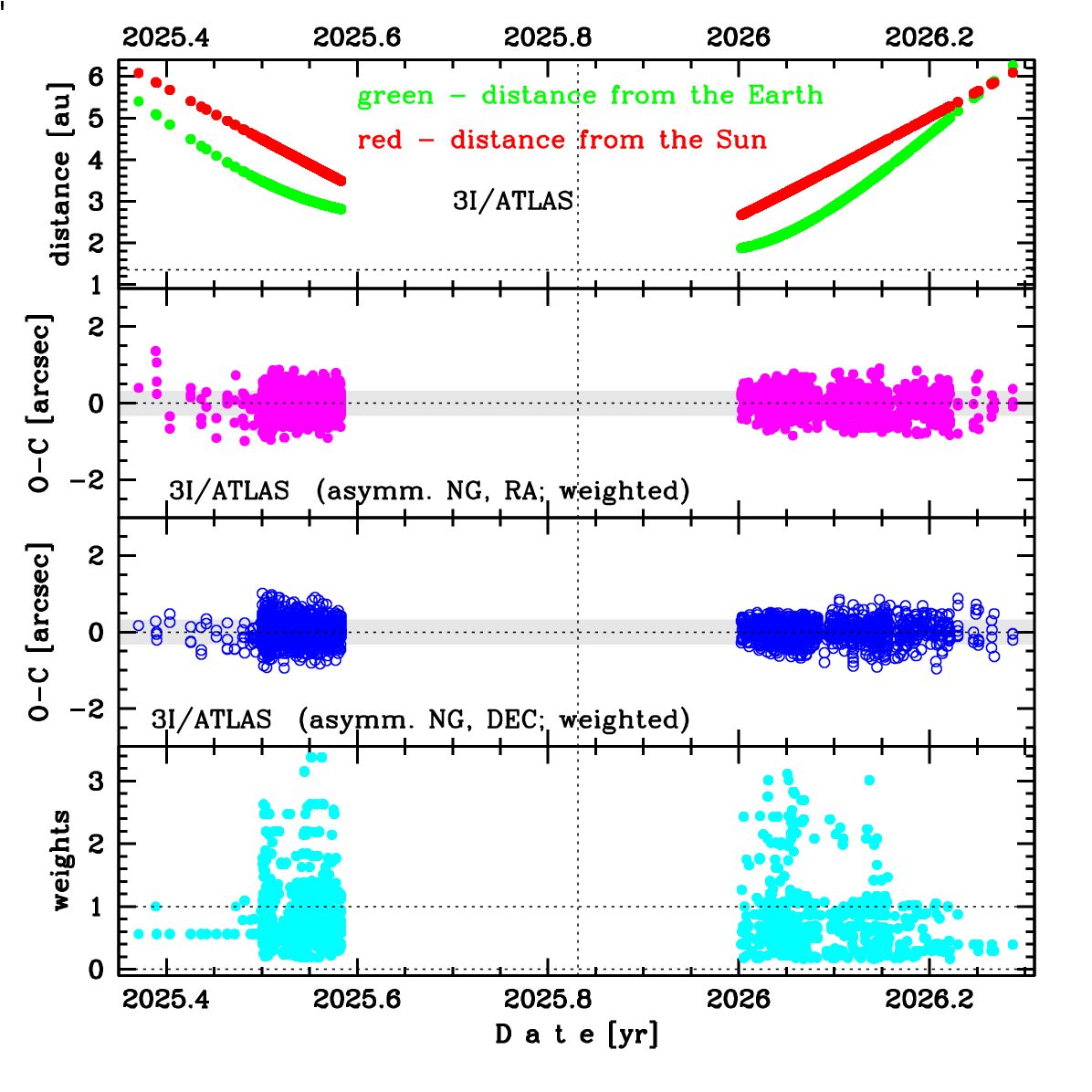}
	\caption{ Residuals for the global solution (data-set3) using GR (left column) and NG-EV (right column) models of motion.  Upper panels: Time distribution of positional observations with corresponding heliocentric (red curve) and geocentric (green curve) distance at which they were taken. The horizontal dotted line shows the perihelion distance for the comet, whereas the vertical dotted line shows the moment of perihelion passage. Middle panels: [O-C] diagram for right ascension and declination, respectively. In the lowest panels, the relative weights are shown. 
  }	\label{fig:3Iglobal-residuals}
\end{figure*}

\section{Global Orbital Solutions}
\label{sec:full_models}

As described above, the pre- and post-perihelion observations were combined into a single data set comprising 4403 astrometric measurements obtained between 2025 May 15 and 2026 April 14. The resulting observational arc spans nearly eleven months and covers large heliocentric distances on both sides of perihelion.

The weights assigned to the observations in data-set3 were inherited from the independent processing of the pre- and post-perihelion data sets and were not recomputed after merging the two arcs. This approach preserves the statistical properties established for the individual data sets and allows a direct comparison between the corresponding orbital solutions.

In particular, the rms values obtained for the separate GR~solutions are 0.256 and 0.250 arcsec for the pre- and post-perihelion arcs, respectively (see Table~\ref{tab:data-sets}). Combining the two weighted data sets would therefore be expected to yield a global rms of approximately 0.26 arcsec, corresponding to the weighted quadratic mean of the rms values obtained for the separate solutions. This would be the natural expectation if a single dynamical model provided an equally good description of both arcs.
Instead, the GR~solution for data-set3 gives rms = 0.328 arcsec, indicating a substantial deterioration of the fit when the two distant arcs are modeled simultaneously. This suggests that a GR~model is unable to provide a consistent description of the motion over this observational interval.

Indeed, the inclusion of NG~acceleration leads to a substantial improvement of the orbital fit. The rms decreases to 0.255 arcsec for the symmetric NG-std  model and further to 0.253 arcsec and 0.254 arcsec for the NG-MV and NG-EV  activity laws, and the differences in rms for the three symmetric NG~models considered are negligible.  Remarkably, this improvement is obtained despite the complete exclusion of observations acquired nearest perihelion, where NG effects would normally be expected to be strongest.

An equally important result is the disappearance of systematic temporal trends in the O$-$C residuals. Such trends are clearly visible for the purely gravitational solution but vanish for all investigated NG models. Figure~\ref{fig:3Iglobal-residuals} presents a representative comparison of the residuals for the GR and asymmetric NG-EV solutions. The simultaneous reduction of the rms and the elimination of systematic residual patterns provide two independent lines of evidence that nongravitational forces are required by the astrometric data.

The statistical properties of the residual distributions provide additional insight into the quality of the solutions. In right ascension, the differences between the GR and NG solutions are relatively modest. The GR~solution already exhibits only moderate departures from normality, with an Anderson--Darling statistic of 2.18 and a D'Agostino--Pearson p-value of 0.15. Thus, although the Anderson--Darling test formally rejects normality, the overall shape of the distribution remains reasonably close to Gaussian.

The NG solutions do not substantially alter this picture. The Anderson--Darling statistics remain at a comparable level, while the D'Agostino--Pearson p-values increase to approximately 0.2--0.5. The residual distributions therefore become slightly more Gaussian than in the GR case, particularly for the asymmetric activity laws. Among all investigated models, the asymmetric formulations yield the highest p-values and the smallest absolute skewness, indicating the most nearly Gaussian residual distributions in right ascension.

It is worth noting that the right-ascension residuals of the global NG solutions are noticeably closer to Gaussian than those obtained for the post-perihelion GR solution alone. Since the post-perihelion observations constitute an integral part of the global data set, this improvement suggests that the inclusion of NG acceleration accounts for at least part of the excess scatter that manifests itself after perihelion.

A much stronger effect is observed in declination. The residual distribution obtained for the GR solution exhibits pronounced departures from normality, with an Anderson--Darling statistic of 9.19 and a D'Agostino--Pearson p-value below $10^{-8}$. This represents the most conspicuous statistical deficiency of the purely gravitational model.

All investigated NG solutions lead to substantial improvement. The mean residuals become nearly zero, the skewness is reduced by approximately an order of magnitude, and the Anderson--Darling statistic decreases from 9.19 to values between about 5.8 and 6.6. Likewise, the D'Agostino--Pearson p-values increase by more than five orders of magnitude (Table~\ref{tab:stat-pre-post}). Although none of the investigated NG models yields a fully Gaussian residual distribution in declination, all move the residual statistics significantly closer to normality. The improvement is remarkably similar for the NG-std, NG-MV, and NG-EV activity laws, again indicating that the existence of NG acceleration is much better constrained than the detailed functional form of its heliocentric dependence.

The overall picture is therefore mixed. On the one hand, the residual distributions remain formally non-Gaussian,
%, particularly in declination, 
demonstrating that a simple model based on constant NG parameters over the entire observational arc does not fully capture the complexity of the comet's motion. On the other hand, every tested NG model produces a substantial and statistically significant improvement relative to the GR~solution.

This result carries an important methodological implication. Residuals that are poorly represented by a purely gravitational orbit should not automatically be regarded as defective observations and removed from the analysis. The selection and weighting procedure adopted here minimizes this risk by relying primarily on the internal residual statistics of individual observing stations rather than on consistency with a specific orbital solution \citep[see][]{bielicki-sitarski:1991,krolikowska-sit-soltan:2009}. Consequently, observations that may contain information on unmodelled dynamical effects are less likely to be discarded. Such residuals may therefore provide useful clues to physical processes not adequately captured by the adopted orbital model.

One possible interpretation of the remaining non-Gaussian residuals, particularly evident in declination, is that the actual NG acceleration varies in magnitude and direction over the observational interval, rather than being adequately described by the constant-parameter model adopted here. The detection of a wobbling high-latitude jet in 3I/ATLAS \citep{SerraRicart2026} and the enhanced sunward CO$_2$ outgassing revealed by JWST observations \citep{Cordiner2025JWST3IATLAS} suggest a heterogeneous and evolving pattern of activity that could produce such variations in the recoil force. The standard NG formalism, however, assumes constant parameters $A_1$, $A_2$, and $A_3$ over the entire observational interval. The remaining departures from Gaussianity may therefore reflect dynamical effects that are not adequately represented by this simplified description. However, the present astrometric data alone do not allow this hypothesis to be tested quantitatively.

\subsection{Asymmetry relative to perihelion}\label{sub:asymmetry}
Interestingly, all asymmetric NG~solutions yield a consistent displacement of the activity maximum relative to perihelion. The derived time shifts  (in days) amount to
\[
DT = -7.75\pm 2.41 
\] 
for the NG-std model yelding rms$=0.255$~arcsec,
\[
DT = -5.88 \pm 2.61  
\]
for the NG-MV model with rms$=0.253$~arcsec, and 
\[
DT = -5.45\pm 2.67 
\]
for the NG-EV model with rms$=0.254$~arcsec. The NG-EV orbital solution is shown in  Appendix~\ref{sec:appA} (Table~\ref{tab:orbital_solutions}) for comparison with the literature.

Within the formal uncertainties, all three determinations are mutually consistent, suggesting that the inferred distant asymmetry is a robust feature of the astrometric data rather than a consequence of the adopted activity law.

It is noteworthy that the derived negative time shifts differ systematically from the positive offsets reported by \cite{Spada2026} and obtained in the JPL solution. \cite{Spada2026} derived a displacement of the activity maximum of $DT=+7.5\pm0.9$ days relative to perihelion, while the corresponding value listed by JPL is $DT=+9.5\pm1.4$ days (Table~\ref{tab:other-solutions}). In both cases, the solutions were based on sublimation models driven by extremely volatile species and were determined using observations covering the entire available data arc, including measurements obtained near perihelion. The JPL solution, however, was based on only a fraction of the currently available astrometric observations. On the other hand, \cite{Eubanks2025} obtained a substantially negative asymmetry parameter, $DT=-34\pm4$ days, using observations extending from the pre-perihelion arc through perihelion to approximately one month after perihelion. Thus, published determinations of the asymmetry parameter span a remarkably broad range, from strongly negative to significantly positive values.

The present solutions, derived exclusively from observations obtained at heliocentric distances larger than 3.49 au before perihelion and 2.67 au after perihelion, consistently indicate a negative displacement of the activity maximum relative to perihelion. Taken together with the previously published positive and strongly negative determinations, this suggests that the degree, and possibly even the sign, of the asymmetry may depend on heliocentric distance and cannot be adequately represented by a single time-shift parameter over the entire orbit.

One possible interpretation is that different volatile species dominate the outgassing at different heliocentric distances. At large distances from the Sun, where the present data are concentrated, the activity may be controlled primarily by highly volatile species such as CO$_2$ and CO, whereas near perihelion (q=1.356 au) water-ice sublimation may contribute significantly to the nongravitational acceleration. In such a scenario, the commonly adopted assumption of a single, fixed asymmetry parameter throughout the entire orbit would represent only an effective approximation to a more complex physical process. 

The apparent discrepancy between the asymmetry parameters derived from different subsets of the astrometric data does not necessarily imply inconsistencies between the corresponding orbital solutions. Rather, it may indicate that a single time-shifted activity law is insufficient to represent the actual heliocentric evolution of the NG~acceleration. In this sense, the fitted value of DT should be interpreted as an effective parameter describing the particular portion of the orbit sampled by the observations rather than as a unique physical characteristic of the comet.

\subsection{Comparison of NG~parameters with the \citet{Spada2026} 'ng2'-solution cited in Table~\ref{tab:other-solutions}}\label{sub:ng2-comp}

Although the two asymmetric NG solutions were derived from substantially different astrometric data sets -- the present solution NG-EV using only distant observations with all data near perihelion excluded, and the 'ng2'-solution of \cite{Spada2026} (Table~\ref{tab:other-solutions}) based on nearly the entire available astrometric arc (data arc: 2025 May 8 -- 2026 Feb 19) -- they exhibit an overall consistency. The transverse and normal components agree within 0.34 and 1.51 combined standard deviations, respectively. In contrast, the radial component differs by about 4.2 combined standard deviations, while the inferred time shift changes from $DT=-5.45\pm 2.67$~d to $+7.49 \pm 0.92$~d, corresponding to a difference of about 4.6 combined standard deviations. This comparison illustrates that the inferred transverse and normal components of NG acceleration are robust, whereas the partitioning between the radial component and the adopted asymmetry parameter remains sensitive to the observational data set and the adopted data treatment. In practice, $A_1$ and $DT$ are often the most strongly coupled parameters in asymmetric NG~solutions, as shifting the activity maximum relative to perihelion primarily affects the radial component of the NG~acceleration and can even change the sign of $A_1$ (see the \cite{Eubanks2025} solution listed in Table~\ref{tab:other-solutions}).

\section{How Well Is the Activity Law Constrained?}
\label{sec:other_ng_models}

While the existence of NG acceleration is strongly supported by the astrometric data, its physical origin is less straightforward to establish. In this section we examine whether the available observations are capable of discriminating between different forms of the activity law, in particular between the standard water-ice sublimation law and the NG-MV and NG-EV laws motivated by the observations of 3I discussed in Section~\ref{sec:intr}.

\subsection{Implications for Extremely Volatile Activity Laws}
\label{subsec:NG-EV}

Figure~\ref{fig:g_r_plot} illustrates the three activity laws considered in this study. Over the heliocentric distance range sampled by the observations (indicated by the blue and red vertical bands), the adopted NG-MV and NG-EV sublimation laws exhibit very similar heliocentric dependences, whereas the standard sublimation law differs substantially from both. The observational arcs extend from approximately 6.1~au to 3.5~au before perihelion and from 2.7~au to 6.1~au after perihelion, thus covering the part of $g(r)$ beyond the characteristic turnover distance of the standard water-ice law near 2.8~au. However, one might expect the astrometry to distinguish between the standard water-ice prescription and the significantly flatter two remaining $g(r)$-like laws, while providing only limited discrimination between the NG-MV and NG-EV formulations themselves.

At the same time, the actual NG~acceleration may result from the combined contribution of different volatile species whose relative importance changes with heliocentric distance. An effective heliocentric dependence resulting from such a combination need not coincide exactly with any single sublimation law. In particular, over a limited heliocentric-distance range, the combined activity of different volatile species may produce an effective dependence that is approximately described by an inverse-square law. In this context, the NG-EV law considered here provides an example of an activity law whose heliocentric dependence may resemble an inverse-square dependence over the relevant range.
	
The exclusion of observations near perihelion should also be considered in this context. It was introduced to avoid the rapidly changing and potentially more complex activity behaviour close to perihelion, rather than to remove the dynamical effect of NG~activity at small heliocentric distances. Although the observations used here do not directly sample this region, the post-perihelion observations refer to an orbit that has already been modified by NG~acceleration also acting around perihelion, where the NG acceleration was expected to be strongly influenced by water-ice sublimation, represented here by the NG-std model. Consequently, the combined distant pre- and post-perihelion data contain information both on the NG~acceleration acting over the observed distant parts of the orbit and on the accumulated dynamical effect of activity at smaller heliocentric distances.
	
Finally, all three NG solutions produce virtually indistinguishable orbital fits. The rms residual is only about 0.001--0.002 arcsec larger for the standard NG law than for the NG-MV and NG-EV models; however, such small differences cannot be regarded as statistically decisive. Furthermore, all three solutions yield comparable residual distributions and are equally successful in removing the systematic trends present in the purely gravitational solution. The resulting orbital elements and NG parameters are likewise mutually consistent within their uncertainties.

These results indicate that, although the distant astrometric observations provide strong evidence for the existence of NG acceleration, they offer only limited leverage for distinguishing the heliocentric dependences prescribed by the three tested models. The effective acceleration may reflect a combination of activity regimes operating over different heliocentric distance ranges, further complicating the interpretation of the inferred acceleration law. Consequently, the choice of the NG-EV activity law remains a physically motivated interpretation supported by independent observational evidence, rather than a requirement imposed by the astrometric data themselves.

More generally, the present observations do not allow us to determine whether the heliocentric dependence of the NG acceleration follows one of the tested prescriptions or a more complex activity law. In particular, a law featuring a characteristic turnover (“knee”) at $r_0$ cannot be ruled out, as the absence of astrometric observations at heliocentric distances greater than 6.1~au prevents us from adequately constraining the behaviour of the acceleration in this regime.

In summary, the available distant astrometry strongly constrains the existence of NG acceleration but does not, within the models tested here, uniquely determine its heliocentric dependence.

\section{Effect of restricting the post-perihelion data to $r_{\rm h}>3.27$~au}\label{sec:test}

In the above analysis, we used a data set outside a clearly asymmetric heliocentric-distance coverage around perihelion, spanning 3.49~au -- 1.37~au (perihelion) -- 2.67~au. This choice was motivated by the aim of retaining a comparable number of observations before and after perihelion. However, such a selection results in an asymmetric heliocentric-distance coverage of the pre- and post-perihelion data. This raises the question of how the derived orbital solutions and, in particular, the inferred time shift $DT$ are affected if the post-perihelion observations are restricted to larger heliocentric distances, making the heliocentric-distance coverage more symmetric with respect to perihelion. The purpose of this section is therefore to test the sensitivity of the derived results to a more symmetric heliocentric-distance selection around perihelion.

We retained the same pre-perihelion data set (data-set1), but restricted the post-perihelion data to heliocentric distances greater than 3.27~au. This reduced the number of post-perihelion observations to 1069, thereby substantially increasing the imbalance between the numbers of observations before and after perihelion; see Table~\ref{tab:setcomp}.

We find that this modification does not improve the statistics describing the deviations of the right ascension or declination distributions from the corresponding Gaussian distributions of the orbit determined using data-set2s, compared with those obtained using data-set2. As in the case of data-set2, reliable NG parameters cannot be determined from these data, as expected given the smaller data span used for the orbital determination. A similar result is obtained for the GR and NG orbits based on data-set3: the $\alpha$ and $\delta$ distributions obtained using data-set3s do not show an improved agreement with Gaussian distributions compared with those obtained using data-set3.

Nevertheless, the results of this test confirm the sensitivity of the derived time shift $DT$ to the heliocentric-distance range of the observations included in the orbital determination. The resulting time shifts ($DT$ in days) are $-4.96\pm2.44$, $-2.87\pm2.69$, and $-2.88\pm2.75$ for the NG-std, NG-MV, and NG-EV models, respectively. Thus, restricting the post-perihelion observations to larger heliocentric distances and thereby making the heliocentric-distance coverage more symmetric leads to a systematic reduction of the time shift relative to perihelion by approximately 2.6--3.0 days compared with the values obtained from the previously adopted data sets. Moreover, for the NG-MV and NG-EV solutions, the uncertainty in $DT$ is comparable to the magnitude of the derived time shift itself, indicating that the corresponding $DT$ values are not significant relative to their uncertainties. It is also worth noting that the transverse component of the NG~acceleration changes sign for the asymmetric NG-EV orbital solution (see the last column of Table~\ref{tab:orbital_solutions} in Appendix~\ref{sec:appA}).

\begin{table}[ht]
	\caption{More limited astrometric data sets used for comparison; rms values are given for the GR~solutions.}
	\label{tab:setcomp}
	\setlength{\tabcolsep}{1.0pt}    \centering
	\begin{tabular}{cccccc}
		\hline \hline
		set name   & data arc & Number        & heliocentric  & rms      \\
		&          & of obs.       & distances [au] & [arcsec] \\
		\hline
		data-set1   & 2025 05 15--2025 07 31  & 2646  & 6.08 -- 3.49    & 0.2558  \\
		data-set2s  & 2026 01 20--2026 04 14  & 1069  & 3.27 -- 6.09    & 0.2325  \\
		data-set3s  & 2025 05 15--2026 04 14  & 3715  & as in data-set1 & 0.3010  \\
		&                              &       & \& data-set2s    &       \\
		\hline
	\end{tabular}
\end{table}

\section{Conclusions}\label{sec:conclusions}

Using astrometric observations obtained outside the interval from $r_h=$3.49 au before perihelion to $r_h=$2.67 au after perihelion, we arrive at the following conclusions for 3I/ATLAS:

\begin{itemize}
	\item The pre-perihelion residuals ($r_h$: 6.1--3.5 au) are close to Gaussian and show no significant temporal trends, indicating that a purely gravitational model provides an adequate description of the motion over this part of the orbit.
	\item The post-perihelion residuals ($r_h$: 2.7--6.1 au) exhibit statistically significant departures from normality despite the absence of obvious temporal trends in the O$-$C residuals, suggesting that additional effects may be present in the data.
	\item Separate analyses of the distant pre- and post-perihelion arcs do not allow statistically meaningful determinations of nongravitational parameters.
	\item When the distant pre- and post-perihelion observations are combined into a single data set, nongravitational acceleration becomes clearly detectable. The rms decreases from 0.33 arcsec for the purely gravitational solution to 0.255--0.254--0.253 arcsec for the NG~solutions, and the systematic O$-$C trends disappear.
	\item All tested NG models significantly improve the fit and produce residual distributions closer to Gaussian, particularly in declination, although statistically significant departures from normality remain for declination. 
	%In right ascension also improvement for NG solutions are obtained, however even for GR solution the distribution does not differ as substantially from Gaussian as in declination.
	\item Asymmetric NG solutions consistently indicate a negative displacement of the activity maximum relative to perihelion. However, comparison with previously published solutions suggests that the inferred asymmetry parameter depends strongly on the portion of the orbit included in the fit and may not represent a unique physical property of the comet.
	\item The astrometric observations do not uniquely constrain the heliocentric dependence of the NG~acceleration. Standard water-ice, NG-MV, and NG-EV activity laws produce nearly indistinguishable orbital fits and residual statistics over the heliocentric distance range covered by the available observations.
\end{itemize}

Based solely on astrometric evidence, the existence of nongravitational acceleration in 3I/ATLAS appears to be considerably better constrained than its physical origin. The remaining departures from Gaussianity may indicate that the actual acceleration field is more complex than represented by the simple empirical models currently used in orbit determination.

\paragraph{\bf Data availability}

The astrometric data underlying this article are publicly available through the Minor Planet Center database. Additional material generated during this study will be shared by the corresponding author upon reasonable request.

\begin{acknowledgements}
	This research has made use of  the positional data of comets provided by the International Astronomical Union’s Minor Planet Center. We would like to thank Man-To Hui for providing a remarkably thorough and detailed report, which helped us improve the clarity and presentation of our results.
\end{acknowledgements}

\bibliographystyle{aa}   
\bibliography{PAD33.bib}

%\bibliographystyle{aa}
%\bibliography{Oort_long_data_arcs_2023jul25}

\clearpage
\onecolumn

\begin{appendix}

\section{Orbital solutions}\label{sec:appA}

\begin{table*}[!ht]
	\caption{Pre-perihelion heliocentric osculating GR orbit and three global asymmetric NG solutions based on NG-EV model of activity; all  based on the distant data arc as described in the paper where the third solution given in the last column is based on smaller set of post-perihelion data than the previous two -- as described in Sect~\ref{sec:test}; Epoch: 2025 11 21 = JD 2461000.5, Equinox=J2000.0. NG parameters are in units of $10^{-8}$au$\cdot$day$^{-2}$, and $DT$ in days. }  
	\label{tab:orbital_solutions}
	\setlength{\tabcolsep}{8.0pt}    \centering
	\begin{tabular}{rrrrr}
		\hline \hline 
		orbital           &    Pre-perihelion  & \multicolumn{3}{c}{Global asymmetric NG solutions} \\
		        elements  &   GR solution      &  used here as basic & dedicated weighting &  less observations \\
		                  &                    &                     & -- for comparison   &  -- for comparison \\
		\hline 
		T        &    20251029.477610557 &  20251029.481181232     &  20251029.481216556 &  20251029.480931123\\
		&	   $\pm$ 0.000684731 &   $\pm$ 0.000129361             &   $\pm$ 0.000136019 &   $\pm$ 0.000131420\\
		q   	 & 	      1.356868048914 &         1.356415837823  &         1.356418884933 &       1.356436258446 \\
		& 	$\pm$ 0.000082396687 & 	 $\pm$ 0.000005401640          &   $\pm$ 0.000005640487 & $\pm$ 0.000006497575\\
		e        &        6.142500852316 &         6.139105781168  &         6.139136551100 &       6.139172701811  \\
		&  $\pm$ 0.000543443374 &   $\pm$ 0.000054499280           &   $\pm$ 0.000057055980 & $\pm$ 0.000056749722 \\
		$\omega$ &      128.010031534    &       128.004589166     &       128.004586723    &     128.005570159   \\
		& 	$\pm$ 0.000746847    &   $\pm$ 0.000204774             &   $\pm$ 0.000212460    & $\pm$ 0.000254462\\  	
		$\Omega$ &      322.164719671    &       322.154308549     &       322.154270020    &     322.154713929  \\
		& 	$\pm$ 0.000852251    &   $\pm$ 0.000090023             &   $\pm$ 0.000093423    & $\pm$ 0.000121736\\ 	
		$i$      &      175.113491556    &       175.113278362     &       175.113271217    &     175.113333278  \\
		& 	$\pm$ 0.000043694    &   $\pm$ 0.000012648             &   $\pm$ 0.000013077    & $\pm$ 0.000016129\\  		
		A$_1$    &      --               &         7.3246          &         7.0864         &         5.9480 \\
		&                       &   $\pm$ 0.4561		           &   $\pm$ 0.4740         &   $\pm$ 0.5767 \\
		A$_2$    &      --               &         0.93806         &         1.0639         &        -2.2272 \\
		& 	                     &   $\pm$ 0.61112	               &   $\pm$ 0.6288         &   $\pm$ 0.8033 \\	
		A$_3$    &      --               &      $-$0.73353         &      $-$0.72710        &      $-$0.81447 \\
		&                       & 	 $\pm$ 0.01443 	               &   $\pm$ 0.01476        &   $\pm$ 0.01914 \\	
		$DT$     &      --               &      $-$5.448           &      $-$6.235          &      $-$2.878  \\
		&                       & 	 $\pm$ 2.674 	               &   $\pm$ 2.933          &   $\pm$ 2.749 \\	
rms      & 0.2558 arcsec  & 0.2541 arcsec & 0.2562 arcsec & 0.2530 arcsec  \\     
		\hline 
	\end{tabular}
\end{table*}

%\FloatBarrier
% teoretycznie nie może przenieść tego co wyżej za ten punkt

Solution in the fourth column of Table~\ref{tab:orbital_solutions} called as dedicated is given only for comparison. Term 'dedicated' means that the procedure of selecting and weigthing was used for the whole distant data (data-set3) together with the asymmetric NG~orbit determination within the NG-EV model.  This results in rms=0.256 arcsec and about 3\% of observations rejected as outliers; compare with 0.254 arcsec and 1.6\% rejected outliers in the case of the basic solution discussed in the main text. Notice that both these NG~solutions are in excellent agreement within combined $1\sigma$ and characterized by similar goodness of fit of the residual distributions in $\alpha$ to Gaussian  and weaker in $\delta$ for the solution with dedicated weighting compared to the basic one.

%($\sigma_{\alpha}=0.262$~arcsec; $\sigma_{\delta}=0.220$~arcsec)

\end{appendix}

\twocolumn

\end{document}